\documentclass[sigconf]{acmart} 

\copyrightyear{2025}
\acmYear{2025}
\setcopyright{cc}
\setcctype{by}
\acmConference[WWW Companion '25]{Companion Proceedings of the ACM Web
Conference 2025}{April 28-May 2, 2025}{Sydney, NSW, Australia}
\acmBooktitle{Companion Proceedings of the ACM Web Conference 2025 (WWW
Companion '25), April 28-May 2, 2025, Sydney, NSW, Australia}
\acmDOI{10.1145/3701716.3715508}
\acmISBN{979-8-4007-1331-6/25/04}

\usepackage{multirow}
\usepackage{enumitem}
\usepackage[ruled,linesnumbered]{algorithm2e}
\usepackage{bm}
\usepackage{xspace}
\newcommand{\model}{\textsf{RALLRec}\xspace}
\usepackage{graphicx}
\usepackage{url}
\usepackage{balance}

\AtBeginDocument{%
  \providecommand\BibTeX{{%
    \normalfont B\kern-0.5em{\scshape i\kern-0.25em b}\kern-0.8em\TeX}}}

\begin{document}

\title[RALLRec]{RALLRec: Improving Retrieval Augmented Large Language Model Recommendation with Representation Learning}


\author{Jian Xu\textsuperscript{$\ast$}}
\affiliation{%
  \institution{Tsinghua University}
  \city{Beijing}
  \country{China}
}
\email{xujian20@mails.tsinghua.edu.cn}

\author{Sichun Luo\textsuperscript{$\ast$}}

\affiliation{%
  \institution{City University of Hong Kong}
  \city{Hong Kong}
  \country{China}
}
\affiliation{%
  \institution{City University of Hong Kong Shenzhen Research Institute}
  \city{Shenzhen}
  \country{China}
}
\affiliation{%
  \institution{
Dongguan University of Technology}
  \city{Dongguan}
  \country{China}
}
\email{sichun.luo@my.cityu.edu.hk}

\author{Xiangyu Chen}
\affiliation{%
  \institution{Tsinghua University}
  \city{Beijing}
  \country{China}
}
\email{xy-c21@mails.tsinghua.edu.cn}

\author{Haoming Huang}
\affiliation{%
  \institution{Alibaba Group}
  \city{Shenzhen}
  \country{China}
}
\email{huanghaoming.hhm@alibaba-inc.com}

\author{Hanxu Hou$^{\dagger}$}
\affiliation{%
  \institution{Dongguan University of Technology}
  \city{Dongguan}
  \country{China}
}
\email{houhanxu@163.com}

\author{Linqi Song$^{\dagger}$}
\thanks{
\parbox{\linewidth}{\textsuperscript{*}Equal Contribution\\$^{\dagger}$Corresponding Author}}
\affiliation{%
  \institution{City University of Hong Kong}
  \city{Hong Kong}
  \country{China}
}
\affiliation{%
  \institution{City University of Hong Kong Shenzhen Research Institute}
  \city{Shenzhen}
  \country{China}
}
\email{linqi.song@cityu.edu.hk}

\begin{abstract}
Large Language Models (LLMs) have been integrated into recommendation systems to enhance user behavior comprehension. The Retrieval Augmented Generation (RAG) technique is further incorporated into these systems to retrieve more relevant items and improve system performance. However, existing RAG methods rely primarily on textual semantics and often fail to incorporate the most relevant items, limiting the effectiveness of the systems.

In this paper, we propose \textbf{R}epresentation learning for retrieval-\textbf{A}ugmented \textbf{L}arge \textbf{L}anguage model \textbf{Rec}ommendation (\model). Specifically, we enhance textual semantics by prompting LLMs to generate more detailed item descriptions, followed by joint representation learning of textual and collaborative semantics, which are extracted by the LLM and recommendation models, respectively. Considering the potential time-varying characteristics of user interest, a simple yet effective reranking method is further introduced to capture the dynamics of user preference. We conducted extensive experiments on three real-world datasets, and the evaluation results validated the effectiveness of our method. Code is made public at \url{https://github.com/JianXu95/RALLRec}.
\end{abstract}

\ccsdesc[500]{Information systems~Recommender systems}

\keywords{Retrieval-augmented generation, Large language model, Recommender system}

\maketitle

\section{Introduction}

Recently, large language models (LLMs) have demonstrated significant potential and have been integrated into recommendation tasks \cite{luo2024recranker,luo2024integrating,luo2024large,wu2024survey}. One promising direction for LLM-based recommendations, referred to as LLMRec, involves directly prompting the LLM to perform recommendation tasks in a text-based format \cite{bao2023tallrec,zhang2023chatgpt}.

However, simply using prompts with recent user history can be suboptimal, as they may contain irrelevant information that distracts the LLMs from the task at hand. To address this challenge, ReLLa \cite{lin2024rella} incorporates a retrieval augmentation technique, which retrieves the most relevant items and includes them in the prompt. This approach aims to improve the understanding of the user profile and improve the performance of recommendations. Furthermore, GPT-FedRec \cite{zeng2024federated} proposes a hybrid Retrieval Augmented Generation mechanism to enhance privacy-preserving recommendations by using both an ID retriever and a text retriever.

\begin{figure}[t]
\centering
\includegraphics[width=0.4\textwidth]{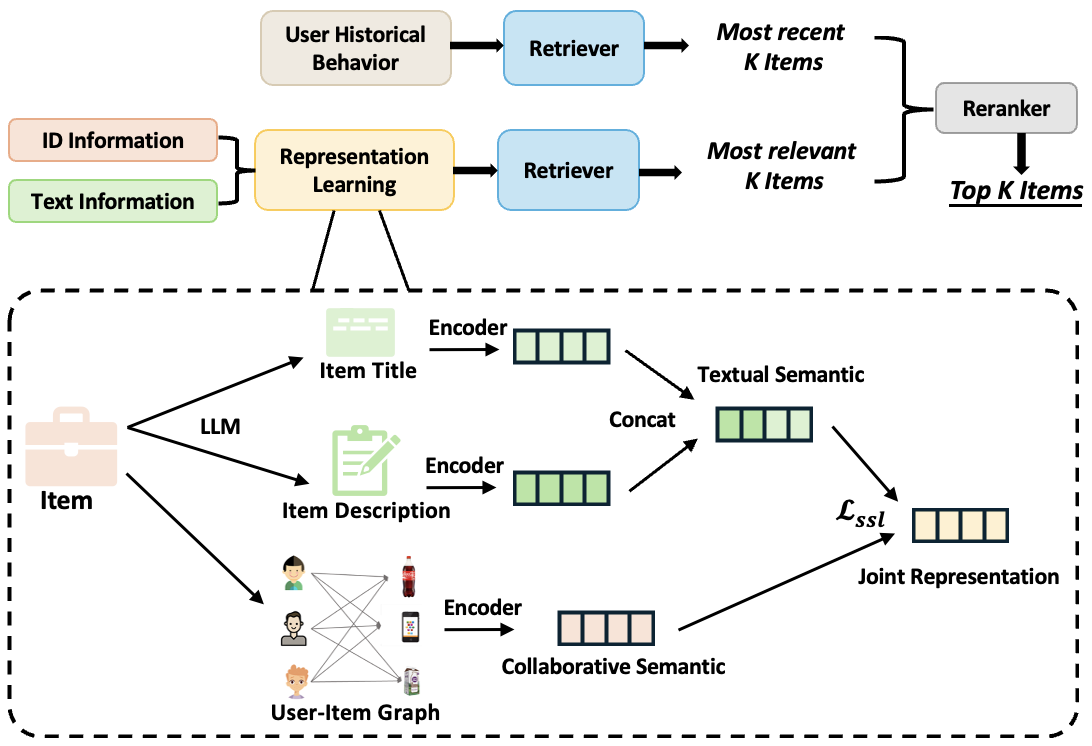}
\vspace{-3ex}
    \caption{\model with embedding, retrieval and reranker.}
\label{fig:pipeline}
\end{figure}

Despite the advancements, current methods have limitations. ReLLa relies primarily on text embeddings for retrieval, which is suboptimal as it overlooks collaborative semantic information from the item side in recommendations. The semantics learned from text are often inadequate as they typically only include titles and limited contextual information. 
GPT-FedRec does not incorporate user's recent interest, and the ID based retriever and text retrieval are in a separate manner, which may not yield the best results.
The integration of text and collaborative information presents challenges as these modalities are not inherently aligned.

In this work, we propose Representation Learning enhanced Retrieval-Augmented Large Language Models for Recommendation (\model). Our objective is to enhance the performance of retrieval-augmented LLM recommendations through improved representation learning.
Specifically, instead of solely relying on abbreviated item titles to extract item representations, we prompt the LLM to generate detailed descriptions for items utilizing its world knowledge. These generated descriptions are used to extract improved item representations. This representation is concatenated with the abbreviated item representation.
Subsequently, we obtain collaborative semantics for items using a recommendation model. This collaborative semantic is aligned with textual semantics through self-supervised learning to produce the final representation. This enhanced representation is used to retrieve items, thereby improving Retrieval-Augmented Large Language Model recommendations.

In a nutshell, our contribution is threefold.
\vspace{-1ex}
\begin{itemize}[left=0em]
    \item We propose \model, which incorporates collaborative information and learns joint representations to retrieve more relevant items, thereby enhancing the retrieval-augmented large language model recommendation.
    

    \item We design a novel reranker that takes into account both the semantic similarity to the target item and the timestamps for boosting the validness of RAG.
    
    \item With extensive explorations of the training and prompting strategies, the experiments reveal several interesting findings and validate the effectiveness of our method.
\end{itemize}




\section{Methodology}

\subsection{Framework Pipeline}

The pipeline of the developed framework is illustrated in Figure~\ref{fig:pipeline}.
The \model encompasses both the retrieval and generation processes.
In the retrieval process, we first learn a joint representation of users and items, allowing us to retrieve the most relevant items in semantic space. These items are then fused with the most recent items by a reranker and incorporated into the prompts. The constructed prompts can be used solely for inference or to train a more effective model through instruction tuning (IT).
For the generation phase, the base LLM responses to the prompt for inference. The base LLM could be standard or customized.

\begin{figure}[t]
\centering
\includegraphics[width=0.43\textwidth]{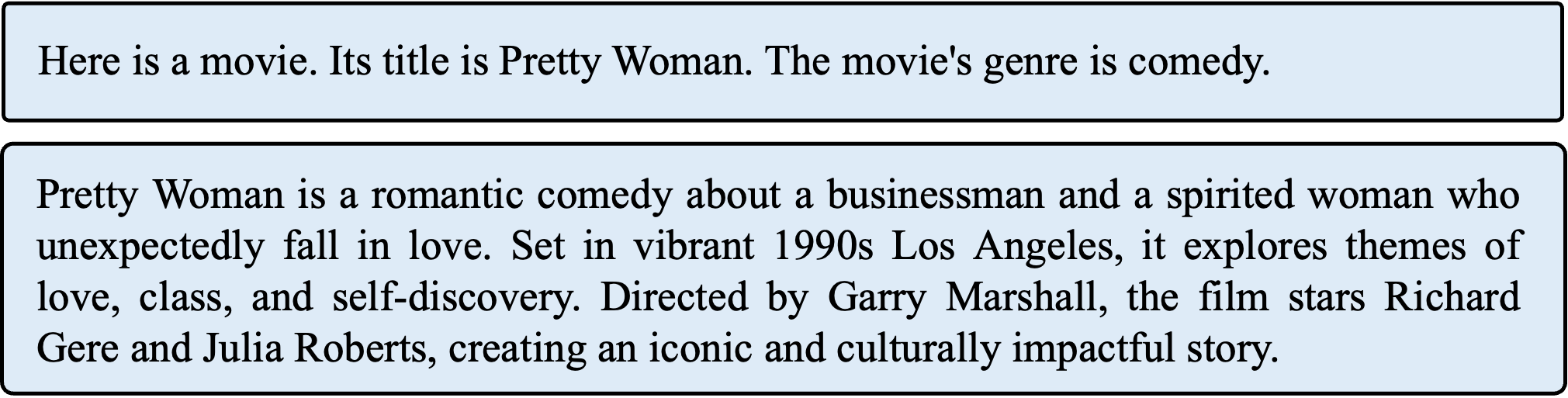}
\vspace{-2ex}
    \caption{Comparison of textual descriptions with fixed template (upper) and automatic generation (blow).}
\label{fig:text_desc}
\vspace{-3ex}
\end{figure}

\subsection{Representation Learning}
To learn better item embeddings \footnote{We interchangeably use the representation and embedding to denote the extracted item feature considering the habits in deep learning and information retrieval domains.} for reliable retrieval, we propose to integrate both the text embedding from textual description and collaborative embedding from user-item interaction, as well as the joint representation through self-supervised training.

\subsubsection{Textual Representation Learning}


In previous work \cite{lin2024rella}, only the fixed text template with basic information such as item title was utilized to extract textual information. However, we argue that relying solely on the fixed text format is inadequate, as it may not capture sufficient semantic depth, e.g., two distinct and irrelevant items may probably have similar names. To enhance this, we take advantage of the LLMs to generate a more comprehensive and detailed description containing the key attributes of the item (e.g., Figure~\ref{fig:text_desc}), which can be denoted as
\begin{equation}
    t^{i}_{\text{desc}} = \text{LLM}(b^{i}|p),
\end{equation}
where $b^{i}$ is the basic information of the $i$-th item and the $p$ is the template for prompting the LLMs.
Subsequently, we derive textual embeddings by feeding the text into LLMs and taking the hidden representation as in \cite{lin2024rella}, represented as 
\begin{equation}
    \mathbf{e}_{\text{desc}}^i = {{\text{LLM}}}_{emb}(t^{i}_{\text{desc}}).
    \label{eq:llm_emb}
\end{equation}
Since the plain embedding of item title $e_{\text{title}}^i$ could also be useful, we aim to directly concatenate these two kinds of embeddings to obtain the final textual representations, denoted by
\begin{equation}
    \mathbf{e}_{\text{text}}^i = [\mathbf{e}_{\text{title}}^i \| \mathbf{e}_{\text{desc}}^i].
    \label{eq:llm_emb}
\end{equation}
It is worth noting that those textual embeddings are reusable after being extracted and they already contain affinity information attributed to the rich knowledge of LLMs.


\subsubsection{Collaborative Representation Learning}

A notable shortcoming of previous LLM-based approaches is their failure to incorporate collaborative information, which is directed learned from the user-item interaction records and thus can be complementary to the text embeddings. To this end, we utilize conventional recommendation models to extract collaborative semantics, denoted as
\begin{equation}
    \{\mathbf{e}_{\text{colla}}^i\}_{i=1}^{n} = \text{RecModel}(\{(u, i)\in \mathcal{V}\}),
\end{equation}
where $n$ is the total number of items and $\mathcal{V}$ is the interaction history.

\begin{table*}[!t]
\caption{
The performance of different models in default settings. The best results are highlighted in boldface. 
The symbol $\ast$ indicates statistically significant improvement of \model over the best baseline
with $p$-value < 0.01.
}
\vspace{-8pt}
\label{tab:zero & few shot performance}
\resizebox{0.95\textwidth}{!}{
\renewcommand\arraystretch{1.0}
\begin{tabular}{c|c|ccc|ccc|ccc}
\toprule

\multicolumn{2}{c|}{\multirow{2}{*}{Model}} & \multicolumn{3}{c|}{BookCrossing} & \multicolumn{3}{c|}{MovieLens} & \multicolumn{3}{c}{Amazon} \\ 
\multicolumn{2}{c|}{} & AUC $\uparrow$  & Log Loss $\downarrow$& ACC $\uparrow$& AUC $\uparrow$ & Log Loss $\downarrow$& ACC $\uparrow$& AUC $\uparrow$ & Log Loss $\downarrow$& ACC $\uparrow$\\ 
   \hline 
   
\multicolumn{1}{c|}{\multirow{4}{*}{ID-based}} & DeepFM & 0.5480&0.8521&0.5212& 0.7184&0.6205&0.6636 &0.6419&0.8281&0.7760\\
\multicolumn{1}{c|}{\multirow{4}{*}{}} & xDeepFM & 0.5541&0.9088&0.5304& 0.7199&0.6210&0.6696&0.6395&0.8055&0.7711 \\
\multicolumn{1}{c|}{\multirow{4}{*}{}} & DCN  & 0.5532&0.9356&0.5189 & 0.7212&0.6164&0.6681&0.6369&0.7873&0.7744 \\
\multicolumn{1}{c|}{\multirow{4}{*}{}} & AutoInt  & 0.5478&0.9854&0.5246& 0.7138&0.6224&0.6613 &0.6424&0.7640&0.7543\\
   \hline

\multicolumn{1}{c|}{\multirow{3}{*}{LLM-based}} & Llama3.1 & 0.5894 & 0.6839 & 0.5418 & 0.5865 & 0.6853 & 0.5591 & 0.7025 & 0.7305 & 0.4719 \\ 
\multicolumn{1}{c|}{\multirow{4}{*}{}} & ReLLa & 0.7125 & 0.6458 & 0.6368 & 0.7524 & 0.6182 & 0.6804 & 0.8401 & 0.5074 & 0.8224  \\ 
\multicolumn{1}{c|}{\multirow{4}{*}{}} & Hybrid-Score & 0.7096 & 0.6409 & 0.6334 & 0.7646 & 0.6149 & 0.6843 & 0.8405 & 0.5065 & 0.8256 \\
\hline

\multicolumn{1}{c|}{\multirow{2}{*}{Ours}} & \model & \textbf{0.7151$^*$} & \textbf{0.6359$^*$} & \textbf{0.6483$^*$} & \textbf{0.7772$^*$} & \textbf{0.6102$^*$} & \textbf{0.6904$^*$} & \textbf{0.8463$^*$} & \textbf{0.4914$^*$} & \textbf{0.8280$^*$} \\ 
\multicolumn{1}{c|}{\multirow{2}{*}{}} 
& \textit{p-value} & \textit{8.69e-4} & \textit{2.35e-3} & \textit{1.22e-3} & \textit{3.00e-6} & \textit{2.05e-5} & \textit{2.58e-3} & \textit{1.39e-4} & \textit{1.96e-5} & \textit{3.88e-3}\\ 
  
   \bottomrule          
\end{tabular}
\vspace{-5ex}
}
\end{table*}

\subsubsection{Joint Representation Learning}

A straightforward approach for integrating above two representations is to directly concatenate the textual and collaborative representations. However, since these representations may not be on the same dimension and scale, this might not be the best choice. Inspired by the success of contrastive learning in aligning different views in recommendations \cite{zou2022multi}, we employ a self-supervised learning technique to effectively align the textual and collaborative representations.
Specifically, we adopt a simple two-layer MLP as the projector for mapping the original text embedding space into a lower feature space and use the following self-supervised training objective
\begin{equation} \footnotesize
    \mathcal{L}_{ssl}^{}=-\mathbb{E} \left\{\log \left[\frac{f\left(\mathbf{e}_{\text{text}}^i, \mathbf{e}_{\text{colla}}^i\right)}{\sum_{v \in \mathcal{V}} f\left(\mathbf{e}_{\text{text}}^i, \mathbf{e}_{\text{colla}}^v\right)}\right] + \log \left[ \frac{f\left(\mathbf{e}_{\text{colla}}^i, \mathbf{e}_{\text{text}}^i\right)}{\sum_{v \in \mathcal{V}} f\left(\mathbf{e}_{\text{colla}}^i, \mathbf{e}_{\text{text}}^v\right)}\right]\right\},
    \label{eq:loss_ssl}
\end{equation}
where $f\left(\mathbf{e}_{\text{text}}^i, \mathbf{e}_{\text{colla}}^v\right)=exp(sim(\text{MLP}(\mathbf{e}_{\text{text}}^i), \mathbf{e}_{\text{colla}}^v))$ and $sim(\cdot)$ is the cosine similarity function. After the joint representation learning, we can get the aligned embedding for each item $i$ as
\begin{equation}
    \mathbf{e}_{\text{ssl}}^i = \text{MLP}(\mathbf{e}_{\text{text}}^i).
\end{equation}

\subsubsection{Embedding Mixture}
Instead of retrieval using different embeddings separately, we find that integrating those embeddings before retrieval can present better performance, therefore we directly concat them after magnitude normalization
\begin{equation}
    \mathbf{e}_{\text{item}} = [\mathbf{\bar{e}}_{\text{text}}||\mathbf{\bar{e}}_{\text{colla}}||\mathbf{\bar{e}}_{\text{ssl}}],
\end{equation}
where $\mathbf{\bar{e}} := \mathbf{{e}}/\|\mathbf{{e}}\|$. With the final item embeddings, we can retrieve the most relevant items to the target item by simply comparing the dot-production for downstream recommendation tasks.




\subsection{Prompt Construction}
To form a prompt message that LLMs can understand, we use a similar template as in \cite{lin2024rella} by filling the user profile, listing the relevant behavior history and instructing the model to give a prediction. We also observed that the pre-trained base LLMs may perform poorly in instruction following. Therefore, we collect a small amount of training data for instruction tuning, where the prompts are constructed with similarity-based retrieval and a \textit{data augmentation} technique is also employed by re-arranging the retrieved sequence according to the timestamp to reduce the impact of item order.


\subsection{Reranker}
Since we can retrieve the most recent $K$ items as well as the most relevant $K$ items, relying solely on one of them may not be the optimal choice. During the inference stage, we further innovatively design a reranker to merge these two different channels. The reranker can be either learning-based or rule-based; in this case, we utilize a heuristic rule-based reranker. For each item, we assign a channel score $\text{S}_{c}$ and a position score $\text{S}_{pos}$. We assign the channel score as $\alpha$ and $(1-\alpha)$ for embedding-based and time-based channel, respectively. The position score is inversely proportional to the position in the original sequence, i.e., $\{1, \frac{1}{2^{\beta}}, ..., \frac{1}{K^{\beta}}\}$. The hyper-parameters $\alpha$ and $\beta$ are tunable. The total score for each item is calculated as the production of these two scores 
\begin{equation}
    \text{Score}^i = \text{S}^{i}_{c} * \text{S}^{i}_{pos}.
\end{equation}
By taking the items with top-$K$ scores, we can obtain a refined retrieval result to maximize the prediction performance.


\section{Experiment}



\subsection{Dataset}
In this paper, we focus on the click-through rate (CTR) prediction \cite{lin2024rella}. 
We utilize three widely used public datasets: BookCrossing \cite{ziegler2005improving}, MovieLens \cite{harper2015movielens},  and Amazon \cite{ni2019justifying}.
For the MovieLens dataset, we select the MovieLens-1M subset, and for the Amazon dataset, we focus on the Movies \& TV subset. We apply the 5-core strategy to filter out long-tailed users/items with less than 5 records. Some statistics are shown in Table~\ref{tab:datasets}. Similar to ReLLa \cite{lin2024rella}, we collect the user history sequence before the latest item and the ratings to construct the prompting message and ground-truth.

\begin{table}[!t]
    \caption{The dataset statistics.}
    \vspace{-8pt}
    \centering
    \resizebox{0.45\textwidth}{!}{
    \renewcommand\arraystretch{1.1}
    \begin{tabular}{c|cccccc}
    \toprule
     Dataset   & \#Users & \#Items & \#Samples & \#Fields & \#Features \\ 
     \midrule
     BookCrossing  & 8,723 & 3,547 & 227,735 & 10 & 14,279 \\
     MovieLens & 6,040 & 3,952 & 970,009 & 9 & 15,905 \\
     Amazon & 14,386 & 5,000 & 141,829 & 6 & 22,387 \\ 
     \bottomrule
    \end{tabular}
    }
    \vspace{-5pt}
    \label{tab:datasets}
\end{table}

\begin{table*}[t]
\caption{ The performance of different variants of \model. We remove different components of \model to evaluate the contribution of each part to the model. 
The best results are highlighted in boldface.
}
\vspace{-6pt}
\label{tab:ablation_train}
\resizebox{0.88\textwidth}{!}{
\renewcommand\arraystretch{1.0}
\begin{tabular}{r|ccc|ccc|ccc}
\toprule
\multicolumn{1}{c|}{\multirow{2}{*}{Model Variant}} & \multicolumn{3}{c|}{BookCrossing} & \multicolumn{3}{c|}{MovieLens} & \multicolumn{3}{c}{Amazon} \\ 
\multicolumn{1}{c|}{} & AUC $\uparrow$  & Log Loss $\downarrow$& ACC $\uparrow$ & AUC $\uparrow$  & Log Loss $\downarrow$& ACC $\uparrow$ & AUC $\uparrow$  & Log Loss $\downarrow$& ACC $\uparrow$ \\ 
\hline 
\model (Ours) & \textbf{0.7151} & \textbf{0.6359} & \textbf{0.6483} & \textbf{0.7772} & \textbf{0.6102} & \textbf{0.6904} & \textbf{0.8463} & \textbf{0.4914} & \textbf{0.8280} \\ 
- \textit{w/o} Data Aug. & 0.7108 & 0.6460 & 0.6460 & 0.7563 & 0.6394 & 0.6452 & 0.8453 & 0.4978 & 0.8226 \\ 
- \textit{w/o} Retrieval & 0.6960 & 0.6425 & 0.6414 & 0.7687 & 0.6199 & 0.6697 & 0.8404 & 0.5037 & 0.8194 \\ 
- \textit{w/o} IT & 0.5857 & 0.6860 & 0.5441 & 0.5865 & 0.6853 & 0.5591 & 0.7120 & 0.7272 & 0.4765 \\ 
\bottomrule          
\end{tabular}
}
\end{table*}


\subsection{Baseline}
We compare our approach with baseline methods, which include both ID-based and LLM-based recommendation systems.
For ID-based methods, we select DeepFM \cite{guo2017deepfm}, xDeepFM \cite{lian2018xdeepfm}, DCN \cite{wang2017deep}, and AutoInt \cite{song2019autoint} as our baseline models. We utilize Llama3.1-8B-Instruct \cite{dubey2024llama} as the base model and LightGCN \cite{He2020LightGCN} to learn collaborative embeddings in our comparisons.
For LLM-based methods, we consider ReLLa \cite{lin2024rella} and a Hybrid-Score based retrieval as in \cite{zeng2024federated}. By default, we apply the LoRA method and 8-bit quantization to conduct instruction-tuning as in \cite{lin2024rella} and the maximum length of history is $K=30$. For the reranker in our method, we search the $\alpha$ over $\{\frac{1}{2}, \frac{2}{3}, \frac{4}{5}\}$ and fix $\beta=1$ in the experiments.

\begin{figure}[t]
\centering
\includegraphics[width=0.48\textwidth]{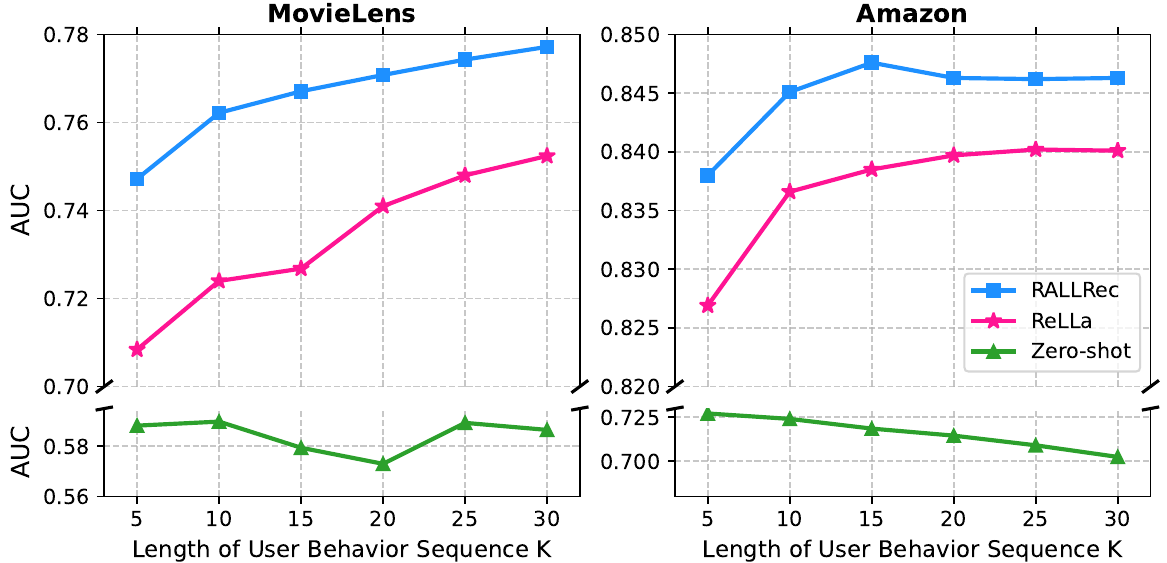}
\vspace{-5ex}
    \caption{The impact of history sequence length K on AUC.}
\label{fig:ablation_K}
\end{figure}

\begin{table*}[t]
\caption{The comparison of different embeddings used for historic behavior retrieval during inference. For fair comparisons, the model is instruction-tuned using the RAG-enhanced training data, while the inference prompt is constructed based on the embedding similarity without re-ranking. The best results are highlighted in boldface.
}
\vspace{-8pt}
\label{tab:ablation_emb}
\resizebox{0.85\textwidth}{!}{
\renewcommand\arraystretch{1.0}
\begin{tabular}{c|ccc|ccc|ccc}
\toprule
\multicolumn{1}{c|}{\multirow{2}{*}{Embedding Variant}} & \multicolumn{3}{c|}{BookCrossing} & \multicolumn{3}{c|}{MovieLens} & \multicolumn{3}{c}{Amazon} \\ 
\multicolumn{1}{c|}{} & AUC $\uparrow$  & Log Loss $\downarrow$& ACC $\uparrow$ & AUC $\uparrow$  & Log Loss $\downarrow$& ACC $\uparrow$ & AUC $\uparrow$  & Log Loss $\downarrow$& ACC $\uparrow$ \\ 
   \hline 
Text-based & 0.7034 & 0.6434 & 0.6426 & 0.7583 & 0.6188 & 0.6798 & 0.8408 & 0.4931 & 0.8222 \\ 
ID-based & 0.7084 & 0.6414 & 0.6357 & 0.7580 & 0.6153 & 0.6867 & 0.8431 & 0.4930 & 0.8244 \\ 
Concat. w/o SSL & 0.7127 & 0.6411 & 0.6391 & 0.7633 & 0.6153 & 0.6828 & 0.8439 & 0.4925 & 0.8244 \\ 
Concat. w/ SSL & \textbf{0.7141} & \textbf{0.6413} & \textbf{0.6471} & \textbf{0.7653} & \textbf{0.6144} & \textbf{0.6850} & \textbf{0.8442} & \textbf{0.4924} & \textbf{0.8269} \\  
\bottomrule          
\end{tabular}
}
\end{table*}

\subsection{Result Analysis}

\textbf{Sequential Behavior Comprehension.} From the numerical results presented in Table~\ref{tab:zero & few shot performance}, several noteworthy observations emerge. Firstly, the vanilla ID-based methods generally underperform LLM-based methods, demonstrating that LLMs can better leverage textual and historical information for preference understanding. 
Secondly, among LLM-based baselines, ReLLa effectively incorporates a retrieval-augmented approach but relies predominantly on simple textual semantics for item retrieval. Hybrid-Score, which considers both ID-based and textual features, also improves over the zero-shot LLM setting (Llama3.1). However, both ReLLa and Hybrid-Score still fail to fully leverage the rich collaborative semantics and the alignment between textual and collaborative embeddings. In contrast, \model consistently achieves the best results across all three datasets, outperforming both ID-based and LLM-based baselines. 
The improvements are statistically significant with $p$-values less than 0.01, emphasizing the robustness of our approach.


\noindent \textbf{Impact of sequence length K.} We change the history length $K$ during the inference stage and collect the final performance in Figure~\ref{fig:ablation_K}. It can be found that as $K$ increases, both \model and ReLLa benefit from longer historical sequences to gain richer insights into user preferences, while the zero-shot LLM baseline suffers from noise and thus does not improve. This phenomenon underscores the importance of carefully selecting and structuring historical user behaviors to assist LLMs in recommendation.
\vspace{-1ex}



\subsection{Ablation Studies}


\subsubsection{Fine-tuning and Data Construction.} We examine the influence of instruction tuning (IT) and data augmentation in Table~\ref{tab:ablation_train}. Removing IT significantly degrades performance, reverting the model to near zero-shot levels, as it struggles to follow the given instructions and task format. Similarly, removing the data augmentation strategy leads to a non-negligible performance drop. This confirms the importance of carefully crafted training data and instruction tuning for aligning the LLM with recommendation objectives.

\vspace{-1ex}
\subsubsection{Reranking and Retrieval Methods.} 
\begin{figure}[t]
\centering
\includegraphics[width=0.49\textwidth]{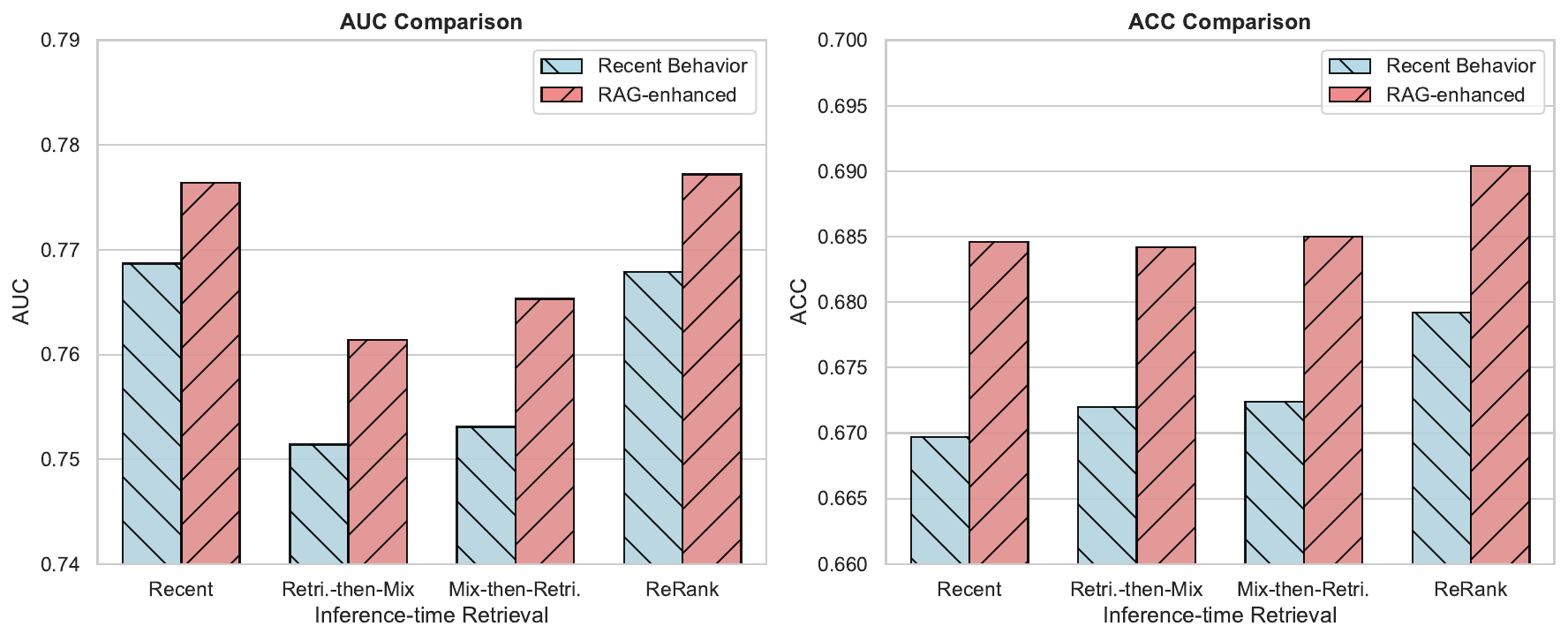}
\vspace{-5ex}
    \caption{Comparison of fine-tuning and inference settings.}
\label{fig:ablation_prompt}
\end{figure}

Figure~\ref{fig:ablation_prompt} compares different retrieval and prompt construction approaches on the MovieLens.
Our reranker, which balances semantic relevance and temporal recency, outperforms both plain recent-history-based prompts and simple hybrid retrieval strategies. These results emphasize the necessity of refining retrieved items through post-processing rather than relying solely on a single retrieval strategy.

\vspace{-1ex}
\subsubsection{Embedding Strategies.} In Table~\ref{tab:ablation_emb}, we contrast various embedding schemes for retrieval. Text-based embeddings provide a decent performance, but they are weaker than the mixture with ID-based embeddings. By aligning textual and collaborative semantics through SSL, we achieve further improvements. 

\vspace{0.5ex}
Overall, the ablation studies confirm that {\textbf{{(i)}}} instruction tuning and data augmentation are essential for aligning the LLM to recommendation tasks, 
{\textbf{{(ii)}}} embedding alignment of textual and collaborative semantics consistently improves retrieval quality, and \textbf{(iii)} a reranking strategy that considers both item relevance and temporal factors enhances the final recommendation performance. Combining these insights, \model presents a robust and effective framework for retrieval-augmented LLM-based recommendation.




\section{Conclusion}
\vspace{-0.5ex}
In this paper, we introduce a new representation learning framework of item embeddings for LLM-based Recommendation (\model), which improves item description generation and enables joint representation learning of textual and collaborative semantics. Experiments on three datasets demonstrate its capability to retrieve relevant items and improve overall performance. 

\begin{acks}
This work was supported in part by the National Natural Science Foundation of China under Grant 62371411, the Research Grants Council of the Hong Kong SAR under Grant GRF 11217823, the Collaborative Research Fund C1042-23GF, and InnoHK initiative, the Government of the HKSAR, Laboratory for AI-Powered Financial Technologies.
\end{acks}

\bibliographystyle{ACM-Reference-Format}
\balance
\bibliography{ref}

\end{document}